\documentclass[aps,prd,reprint,superscriptaddress,floatfix,twocolumn,nofootinbib]{revtex4-2}

\usepackage{color}
\usepackage{amssymb}
\usepackage{graphicx,color}
\usepackage[usenames,dvipsnames]{xcolor}
\usepackage[section]{placeins}
\usepackage{amsmath}
\usepackage{subfigure}
\usepackage[normalem]{ulem}
\usepackage{booktabs}
\usepackage{xcolor}
\usepackage{epsfig,graphics,color,graphicx,amsmath}

\colorlet{darkgreen}{green!50!black}
\colorlet{brightyellow}{yellow!75!red}
\colorlet{orange}{red!50!yellow}
\colorlet{darkblue}{blue!60!black}
\colorlet{darkred}{red!80!black}

\usepackage{soul}

\usepackage{mathtools}
\usepackage{mathrsfs}
\usepackage{bm}
\usepackage{relsize}
\usepackage{indentfirst}

\usepackage{xcolor}

\usepackage{amssymb,amsmath,multirow,epsfig,graphicx,color}

\usepackage{epsfig}
\usepackage{graphicx,amsmath}
\def\be{\begin{eqnarray} &&}
\def\nonu{\nonumber \\ &&}

\def\ee{\end{eqnarray}}
\def\psla{\slash \! \!\! }
\def\Psla{\slash \! \! \!\! }

\begin{document}

\title{The  parton distribution function in a pion with  Minkowskian dynamics}

\author{W. de Paula}
\affiliation{Instituto Tecnol\'ogico de Aeron\'autica,  DCTA,
12228-900 S\~ao Jos\'e dos Campos,~Brazil}
\author{E.~Ydrefors}
\affiliation{Institute of Modern Physics, Chinese Academy of Sciences, Lanzhou 730000, China}
\author{J.H.~Alvarenga Nogueira}
\affiliation{Instituto Tecnol\'ogico de Aeron\'autica,  DCTA,
12228-900 S\~ao Jos\'e dos Campos,~Brazil}
\affiliation{Dipartimento di Fisica, Universit\`a di Roma La Sapienza", P.le A. Moro 2, 00185 Rome, Italy}
\affiliation{
INFN, Sezione di Roma, P.le A. Moro 2, 00185 Rome, Italy}
\author{T.~Frederico}
\affiliation{Instituto Tecnol\'ogico de Aeron\'autica,  DCTA,
12228-900 S\~ao Jos\'e dos Campos,~Brazil}
\author{G. Salm\`e}
\affiliation{
INFN, Sezione di Roma, P.le A. Moro 2, 00185 Rome, Italy}

\date{\today}

\begin{abstract}
The  parton distribution of the pion is obtained for the first time from the solution of a dynamical equation in Minkowski space. The adopted equation is the   homogeneous Bethe-Salpeter one with   a   ladder kernel,   described in
terms of i) constituent quarks and gluons degrees of freedom, and ii) an extended
quark-gluon vertex. The masses of quark and gluon as well as the  interaction-vertex scale have been  chosen in a range suggested by
 lattice QCD  calculations, and calibrated to reproduce
 both pion mass and decay constant.  In addition to the full parton distribution, we have also calculated the contribution from the  light-front valence wave function, corresponding to  the lowest Fock component in the expansion of the pion state. After applying  {an evolution with an effective charge and a LO splitting function}, a detailed and inspiring comparison with both the extracted experimental data ( with and without resummation effects) and  other  recent calculations obtained in different frameworks is presented.
Interestingly, in a wide  region of  longitudinal-momentum fraction, the parton  distribution function receives sizable contributions
  from the higher Fock-components
  of the pion state at the initial scale,   while  {approaching the tail}  the light-front valence  component dominates, as expected. Moreover,  an exponent $\sim 3$ is found suitable for describing the tail at the scale $5.2$ GeV.
\end{abstract}
\maketitle
The pion is a cornerstone for
understanding the visible mass of the universe
within Quantum Chromodynamics (QCD), being the pivotal
Goldstone boson state associated with the dynamical mass generation~
(see, e.g., Ref. \cite{Aguilar:2019teb}).  Dedicated experimental efforts are planned in the close future for  investigating in
detail  the pion, and eventually  to reconstruct  its
3D image   in Minkowski space, by means of high-luminosity
facilities, like
  the  Electron Ion Collider (EIC) in USA~\cite{Accardi:2012qut}, as well the EICc in China~\cite{Anderle:2021wcy}.
 In the perspective to explore dynamical models incorporating non perturbative features of QCD
 and thus able to  gain a reliable  description of  hadrons on the light-cone,  in  this letter we present a    calculation of the parton distribution function (PDF) of the pion and its light-front (LF) valence component, using for the first time a $0^-$ solution of the Bethe-Salpeter  equation (BSE) \cite{BS51} in Minkowski space {(see Ref. \cite{Biernat:2013fka} for a 4D relativistic description of the  pion through the covariant-spectator method)}.  After properly applying an evolution with an effective charge and a LO splitting function (ECLO), namely the suggestion proposed in Ref. \cite{Cui:2021mom} (see also Ref. \cite{Deur:2016tte} for a detailed analysis of the QCD running coupling),   comparisons with data and outcomes from other recent
 calculations,  like continuum QCD \cite{Bednar:2018mtf,Cui:2020tdf}, basis light-front quantization (BLFQ) \cite{Lan:2019rba,Lan:2021wok} and lattice QCD (LQCD) \cite{Alexandrou:2021mmi}
 are illustrated. It is worth noticing that  within our approach,  from the comparison between the full PDF and the LF-valence  contribution (see below) one can quantitatively   assess the phenomenological relevance  of the higher-Fock components
of the pion state.
 
 In order to achieve our goal, we adopt the framework already successfully applied
 to both a  3D investigation of the
 pion \cite{dePaula:2020qna} onto the null-plane and the electromagnetic form factor \cite{Ydrefors:2021dwa} (in a very nice agreement with the data and including  also the asymptotic region). Along with ingredients genuinely belonging to the quantum-field theory realm,
 we use i) the Nakanishi integral
 representation (NIR)~\cite{Nakanishi:1971} of the Bethe-Salpeter (BS) amplitude (see, e.g., Refs \cite{dePaula:2016oct,dePaula:2017ikc} for a
  general introduction to the fermionic case) for  obtaining  solutions of
 the Minkowskian BSE, and ii)   a
 formalism \'a la Mandelstam
\cite{Mandelstam:1955sd} for describing the interaction  between a virtual photon and a bound system, and eventually calculating the PDF.
 
In the ladder approximation,  the bound-state BS  amplitude, $\Phi(k; P)$,  fulfills  the following homogeneous integral equation
\be\label{Eq:BSE}
  \Phi(k; P) = S\bigl(k+\tfrac{P}{2}\bigr) \int \frac{d^4k'}{(2\pi)^4}S^{\mu\nu}(q)\Gamma_{\mu}(q)
 \nonu \times ~ \Phi(k';P)\widehat\Gamma_{\nu}(q)S\bigl(k-\tfrac{P}{2}\bigr)~,
\ee
where  $P$ is the pion 4-momentum, with $P^2 = M^2$, $k=(p_q - p_{\bar q})/2$ the relative 4-momentum, with $p_{q(\bar q)}$ the off-shell (anti-) quark momentum, and $q=k-k'$. The quark-gluon vertex, $\Gamma_{\nu}(q)$,  is related to
$\widehat\Gamma_{\nu}(q)$ by $\widehat\Gamma_{\nu}(q)=C~\Gamma_{\nu}(q)
~C^{-1}$,
where  $C=i\gamma^2\gamma^0$  is the  charge-conjugation operator.
In Eq.~\eqref{Eq:BSE},   the fermion propagator,  the gluon
propagator in the Feynman gauge  and
the   extended quark-gluon  vertex (dressed through a simple form factor)
are
\be
S(p) =  ~{i \over \psla{p} - m + i\epsilon}~, \qquad
    S^{\mu\nu}(q) = -i \frac{g^{\mu\nu}}{q^2 - \mu^2 + i\epsilon}~,
    \nonu~
    \Gamma^\mu= i g \frac{\mu^2 - \Lambda^2}{q^2 - \Lambda^2 + i\epsilon}\gamma^\mu~,
\label{def}\ee
where $g$ is the coupling constant, $m$ the fermionic mass,  $\mu$ the exchanged-boson mass
and $\Lambda$  a scale parameter, introduced for
 modeling the  color distribution  at the
 interaction vertex.    Noteworthy, the one-gluon exchange should be
 a viable approximation according to  Ref. \cite{Nogueira:2017pmj}, where the  non-planar diagrams were found largely $N_c$ suppressed in bosonic bound states (with an estimate of their contribution to dynamical observables
  less than 5\% for  $N_c\,=\,3$, even for large binding).
 %

The BS amplitude for a $0^-$ system   reads
\be\label{BS_decomp}
  \Phi(k;P) =  S_1(k;P) \phi_1(k;P)+S_2(k;P) \phi_2(k;P)
  \nonu{}+S_3(k;P) \phi_3(k;P)+S_4(k;P) \phi_4(k;P)~~,
\ee
 where the $\phi_i$'s are scalar functions, and $S_i$'s are
 Dirac structures  given by~\cite{Lwel,Carbonell:2010zw}
\be
    S_1(k;P) = \gamma_5, \,\, S_2(k;P) = \frac{\Psla{P}}{M}{\gamma_5},
    \nonu
    S_3(k;P) = \frac{k\cdot P}{M^3}\Psla{P}\gamma_5 -
    \frac{1}{M}\psla{k}\gamma_5,
    \nonu S_4(k;P) =
    \frac{i}{M^2}\sigma^{\mu\nu}P_\mu k_\nu \gamma_5\, .
\ee
 The anti-commutation rules of the fermionic fields
 impose that   the functions  $\phi_i$  are
 even  for $i=1,2,4$, under the change
 $k \rightarrow - k$, and odd for $i=3$.

The  scalar functions $\phi_i(k;P)$ in \eqref{BS_decomp} can be written in terms of the NIR as follows
\be
  \label{Eq:NIR}
  \phi_i(k; P) = \int_{-1}^1 dz' \int_0^\infty d\gamma'
  \nonu \times
  \frac{g_i(\gamma',z';\kappa^2)}{[k^2 + z'(P\cdot k) - \gamma' - \kappa^2 + i \epsilon]^3},
\ee
where  $\kappa^2 = m^2 -{M^2/4}$, and  $g_i(\gamma',z';\kappa^2)$ are the
 Nakanishi weight
functions (NWFs), that   are real and
 assumed   to be unique, following    the uniqueness theorem from Ref.~\cite{Nakanishi:1971}.
 Remarkably,
all the  dynamical information one is able to  include  in the BS interaction kernel are non perturbatively  embedded in the NWFs, once the suitable integral equation is solved.

By inserting Eqs.~\eqref{BS_decomp} and \eqref{Eq:NIR} in the BSE,
Eq. \eqref{Eq:BSE}, and then applying a LF projection, i.e. integrating over $k^-=k^0-k^3$ ,
one can formally transform the BSE
into a
 coupled system of integral equations for the NWFs (see details in
Ref.~\cite{dePaula:2017ikc}), that eventually becomes a generalized eigenvalue problem (GEVP). To carry out the numerical evaluation,
the  range of variability  of the constituent quark
  and gluon  effective masses, as well as the scale parameter $\Lambda\sim
\Lambda_{QCD}$ have been chosen as
 suggested  by LQCD results (see, e.g., Refs. \cite{DuPRD14,Rojas2013,Oliveira:2020yac}), as discussed in detail in Ref. \cite{dePaula:2020qna}. In particular,
 by using  i)  $m=255$ MeV,
 ii) $\mu= 637.5$ MeV and iii) $\Lambda= 306$ MeV (the three values correspond to the
set VIII in~\cite{dePaula:2020qna}), one is able
  to reproduce the pion mass
 $M=140$ MeV and the PDG estimation of the decay
 constant $f^{\text{PDG}}_{\pi^-}=$130.50(1)(3)(13)~MeV~\cite{Zyla:2020zbs}. The coupling constant $g$ in the interaction vertex (see Eq.~\eqref{def}) is  also an outcome of the GEVP,  that yields $g^2/(4\pi)=6.482$.
This value is  in a acceptable (factor $\sim 2$)
 agreement  with  $\alpha_s/\pi$ in the IR domain,   presented in the wide analysis of Ref. \cite{Deur:2016tte}.
%
 
{\it The parton distribution function.}
Once the NWFs are numerically calculated, one  obtains the full BS amplitude through Eqs. \eqref{Eq:NIR} and \eqref{BS_decomp}. After performing the normalization  in the standard way \cite{Lurie} (see also Refs. \cite{dePaula:2020qna,Ydrefors:2021dwa}), one   proceeds to   evaluate  the pion PDF. The starting point is the unpolarized transverse-momentum distribution (uTMD), that  adopting the light-cone gauge, $A_g^+=0$, reads in the frame ${\bf P}_\perp=0$
(see, e.g., Refs. \cite{Barone:2001sp,Fanelli:2016aqc})
\be
 f_1(\gamma, \xi)
 =~{N_c\over 4}\int d\phi_{\hat {\bf k}_\perp}
 \int {dz^-  d{\bf z}_\perp \over 2 (2\pi)^3}e^{i [\xi P^+z^-/2 -
 {\bf k}_\perp\cdot
 {\bf z}_\perp] }
 \nonu
 \times~\langle P|
\bar{\psi}_q (-\frac12z) \gamma^+  \psi_q(\frac12z)
|P \rangle\big|_{ z^+=0}
\nonu
={N_c\over 4}\int d\phi_{\hat {\bf k}_\perp} \int {dp^-_q\over 2 (2\pi)}\int dp^+_q~\delta(p^+_q-\xi P^+)
\nonu{\times~} \int {dz^- dz^+ d{\bf z}_\perp \over 2 (2\pi)^3}e^{i [(p^+_q z^- +p^-_qz^+)/2 -
 {\bf k}_{\perp}\cdot
 {\bf z}_\perp] }
\nonu \times ~ \langle P|
\bar{\psi}_q (-\frac12z) \gamma^+  \psi_q(\frac12z)
|P \rangle,
\label{Eq:uTMD}
\ee
where  
$\gamma=|{\bf k}_\perp|^2$, $\xi=p_q^+/P^+=(k^++P^+/2)/P^+$. The uTMD is normalized to 1 given the normalization of the pion state (see Ref. \cite{Ydrefors:2022}).
Then the PDF is nothing more than  the integral over $\gamma$ of the uTMD. i.e.
\be
u(\xi)= \int_0^\infty d\gamma~ f_1(\gamma, \xi).
\label{Eq:pdf}
\ee
By assuming the charge symmetry (see, e.g., Ref.~\cite{Londergan:2009kj}) and adopting the Mandelstam framework~\cite{Mandelstam:1955sd} (see also Ref.~\cite{Ydrefors:2021dwa} for the pion electromagnetic form factor), that heuristically amounts to use a dressed quark-pion vertex (related to the BS amplitude after multiplying by the fermion propagators),   the expression for the uTMD is given by (see Ref. \cite{Ydrefors:2022})
\be
f_1(\gamma,\xi)= {1 \over (2\pi)^4}~{1 \over 8}
\nonu{\times ~}\int_{-\infty}^\infty dk^+ \delta(k^+ + P^+/2- \xi P^+)
\int_{-\infty}^\infty  dk^-
\int_0^{2\pi} d\phi_{\hat {\bf k}_\perp}
\nonu \times
~\Biggl\{ Tr\Bigl[ S^{-1}(k-P/2) \bar \Phi(k,P)~  {\gamma^+ \over 2} ~\Phi(k,P) \Bigr]
\nonu- Tr\Bigl[ S^{-1}(k + P/2) \Phi(k,P)~  {\gamma^+ \over 2} ~ \bar \Phi(k,P) \Bigr] \Biggr\}\, .
\label{Eq:uTMD1}
\ee
Notice that $f_1$ in Eq.~\eqref{Eq:uTMD1} is automatically normalized to 1, once the BS amplitude is normalized (cfr. Refs.~\cite{dePaula:2020qna} and \cite{Lurie}), and that  the explicit expression of Eq.~\eqref{Eq:uTMD1} in terms of the NWFs  is given in Ref. \cite{Ydrefors:2022}.

In summary, our calculation of the PDF is carried out by using in Eq.~\eqref{Eq:pdf} the result of Eq.~\eqref{Eq:uTMD1} with the BS amplitude evaluated through Eqs.~\eqref{Eq:NIR} and \eqref{BS_decomp}. The different gauges in Eq.~\eqref{Eq:uTMD} and in the BSE kernel (at the present stage) raises the question of the relevance of the Wilson line in Eq.~\eqref{Eq:uTMD}, that reduces to the identity in the light-cone gauge. The non trivial challenge of adopting a gluon propagator in the light-cone gauge  will be faced with elsewhere, but one could reliably surmise a small effect after comparing our result with the one in Ref.~\cite{Bednar:2018mtf}, where a Landau gauge has been adopted (see  Fig. \ref{Fig3} for comparison, { modulo the very sharp differences in the approaches}).

In addition to the full PDF, for a more deep analysis we have calculated the LF valence contribution.  Within the LF quantum-field theory illustrated in Refs.~\cite{YanII,YanIV}, one
defines the creation and annihilation operators for
particles and antiparticles, with arbitrary spin,
onto the null-plane.   Then, the   generic LF Fock state is built and, assuming a tiny mass for the gluon, one can meaningfully expand the hadron wave function (WF) by using the complete Fock basis  and diagonalize the
LF Hamiltonian (see Ref.~\cite{Brodsky:1997de}).
The state with the smallest number of constituents (or with the lowest number of creation operators applied to the vacuum)
is the valence one, and we call  the corresponding amplitude
  {\em LF valence WF}. Notice  that in the literature (see, e.g., Ref. \cite{Cui:2021mom} where a detailed analysis of the issue is presented and a wealth of related  references are given) a different terminology is adopted, by indicating as valence WF the full LF-projected BS amplitude, emphasizing in this way the number of fermionic fields, dressed by QCD interactions, that are present  in the definition of the BS amplitude itself.

The Fock expansion of the pion state is a very useful tool, since { in the  $A^+_g=0$ gauge} one can recover a probabilistic framework, inapplicable to  the BS amplitude. In fact,  by summing up   the square modulus
of  each amplitude present  in the Fock expansion, we obtain 1, if the pion state is normalized. With this in mind, one can write the contribution to the PDF from the  LF-valence WF as follows (see   Ref. \cite{dePaula:2020qna})
\begin{equation}
u_{val}(\xi)=
\int_0^\infty {d\gamma\over (4\pi)^2}~\Bigl[|\psi_{\uparrow\downarrow}(\gamma,z)|^2+|\psi_{\uparrow\uparrow}(\gamma,z)|^2\Bigr],
\label{Eq:uval}\end{equation}
where   $z=1-2\xi$, $\psi_{\uparrow\downarrow}(\gamma,z)$ is the anti-aligned component of the LF-valence WF and  $\psi_{\uparrow\uparrow}(\gamma,z)$ the aligned one (of purely relativistic nature).  The probability of the LF-valence WF reads
\be
P_{val}=\int_{-1}^1  dz\int_0^\infty {d\gamma\over (4\pi)^2}~\Bigl[|\psi_{\uparrow\downarrow}(\gamma,z)|^2+|\psi_{\uparrow\uparrow}(\gamma,z)|^2\Bigr] \, .  \nonu
\ee
In the actual calculation { within the Feynman gauge, one gets $P_{val}=0.7$ for the adopted three parameters   \cite{dePaula:2020qna}, and the remaining  probability  }  indicates
that $30\%$ of the normalization comes from Fock-states $|q\bar q; n~gluons\rangle$ with $n=1,2,\dots \infty$.

{\it Results.}
The full PDF and its LF-valence contribution, obtained from the BSE evaluated through the NIR approach and adopting the previously mentioned input parameters,
are shown in Fig. 1, at the initial scale  $Q_0=360$ MeV. This key value for $Q_0$ is chosen in agreement to the analysis of the running coupling that allows us to assign a hadronic scale from the inflection point of the QCD effective charge  as a function of $Q^2$ (see Refs.~\cite{Cui:2021mom}, where $Q_0=0.330\pm 0.030$ GeV was adopted, and also~\cite{Deur:2016tte}). In particular, the actual value allows one to reproduce the LQCD~\cite{Alexandrou:2020gxs} result for the longitudinal-momentum sum rule (see below).

\begin{figure}[htb]
\begin{center}
\epsfig{figure=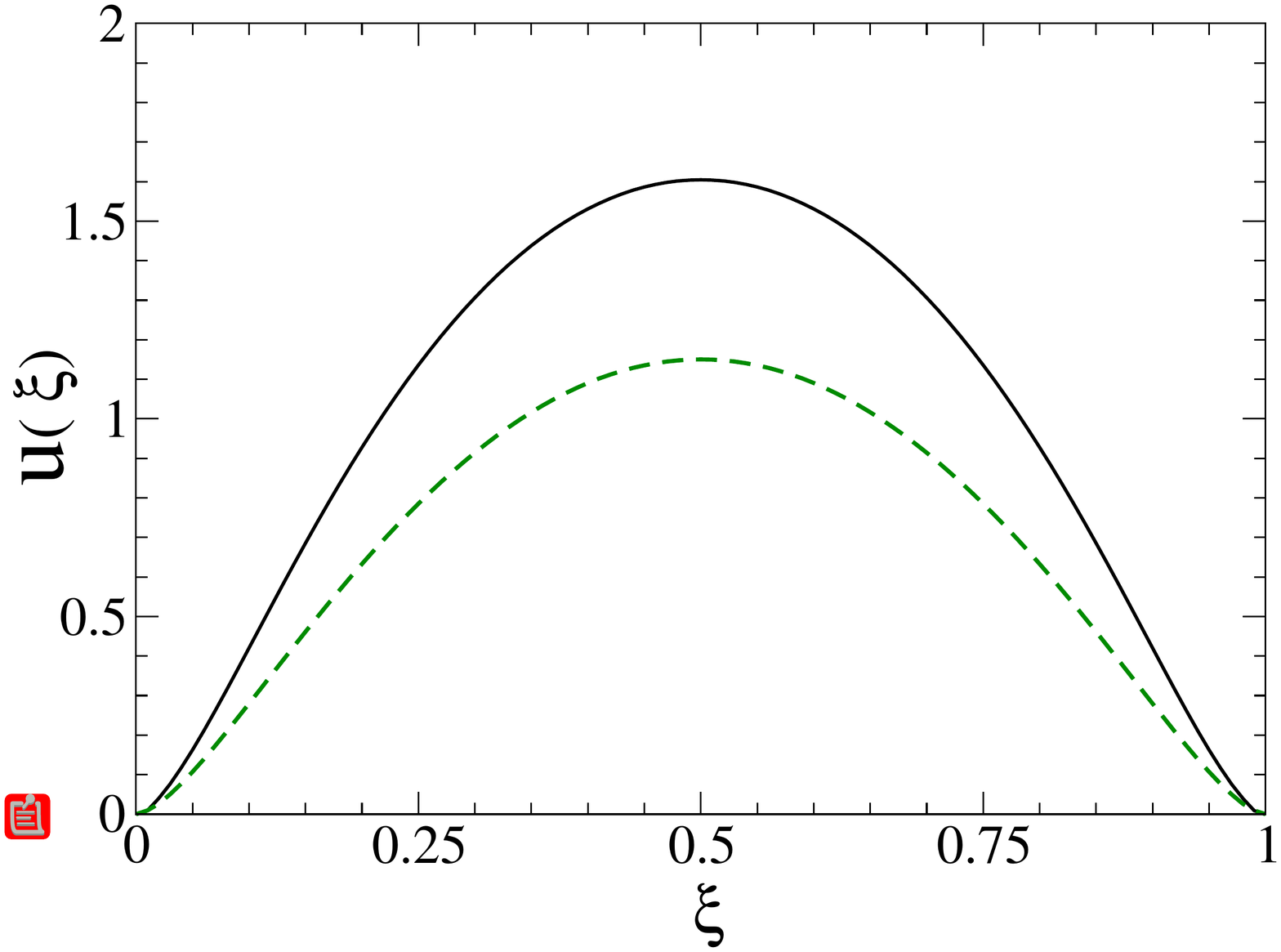,width=9.5cm}
\caption{(Color online). Pion parton distribution function. Solid line: full calculation from Eqs. \eqref{Eq:pdf} and \eqref{Eq:uTMD1}, by using the solution of the BSE, evaluated through the NIR approach and adopting the values of the three input parameters $m= 255$ MeV, $\mu=637.5$ MeV and $\Lambda=306$ MeV (see Ref. \cite{dePaula:2020qna}). Dashed line: LF valence contribution, from the valence component of the Fock expansion
of the pion state, Eq.~\eqref{Eq:uval} (see Ref. \cite{dePaula:2020qna}).}\label{Fig1}
 \end{center}
\end{figure}

Some comments on the results in Fig.~\ref{Fig1} are in order: i) the symmetry of the PDFs, with respect to $\xi=0.5$, is entailed by the charge symmetry, that in turn leads to the expression of the uTDM in the Mandelstam approach given by Eq.~\eqref{Eq:uTMD1}; ii) for $\xi \to 1$, the amplitude of the lowest Fock state generates a contribution that completely saturates $u(\xi)$; iii) while the full PDF is normalized to 1, as it necessarily follows from the standard normalization of the BS amplitude~\cite{Lurie,dePaula:2020qna}, the valence contribution has norm  $P_{val}=0.7$; iv) the $30\%$ depletion is due to the presence of the higher Fock-components in the pion state. {Let us} remind  that the two spin configurations of the quark pair contribute to the valence PDF with different probabilities: $P_{\uparrow \downarrow}=0.57$  and  $P_{\uparrow \uparrow}=0.13$, {so that}  one  remarkably finds a weight of  $\sim 25\%$ from  purely relativistic effects carried by the aligned component (see Ref.~\cite{dePaula:2020qna} for more details).  Finally, at the initial scale, the exponent of $(1-\xi)^{\eta_0}$ for $\xi\to 1$ for the  full PDF is  $\eta_0=1.4$.
\begin{figure}[htb]
\begin{center}
\epsfig{figure=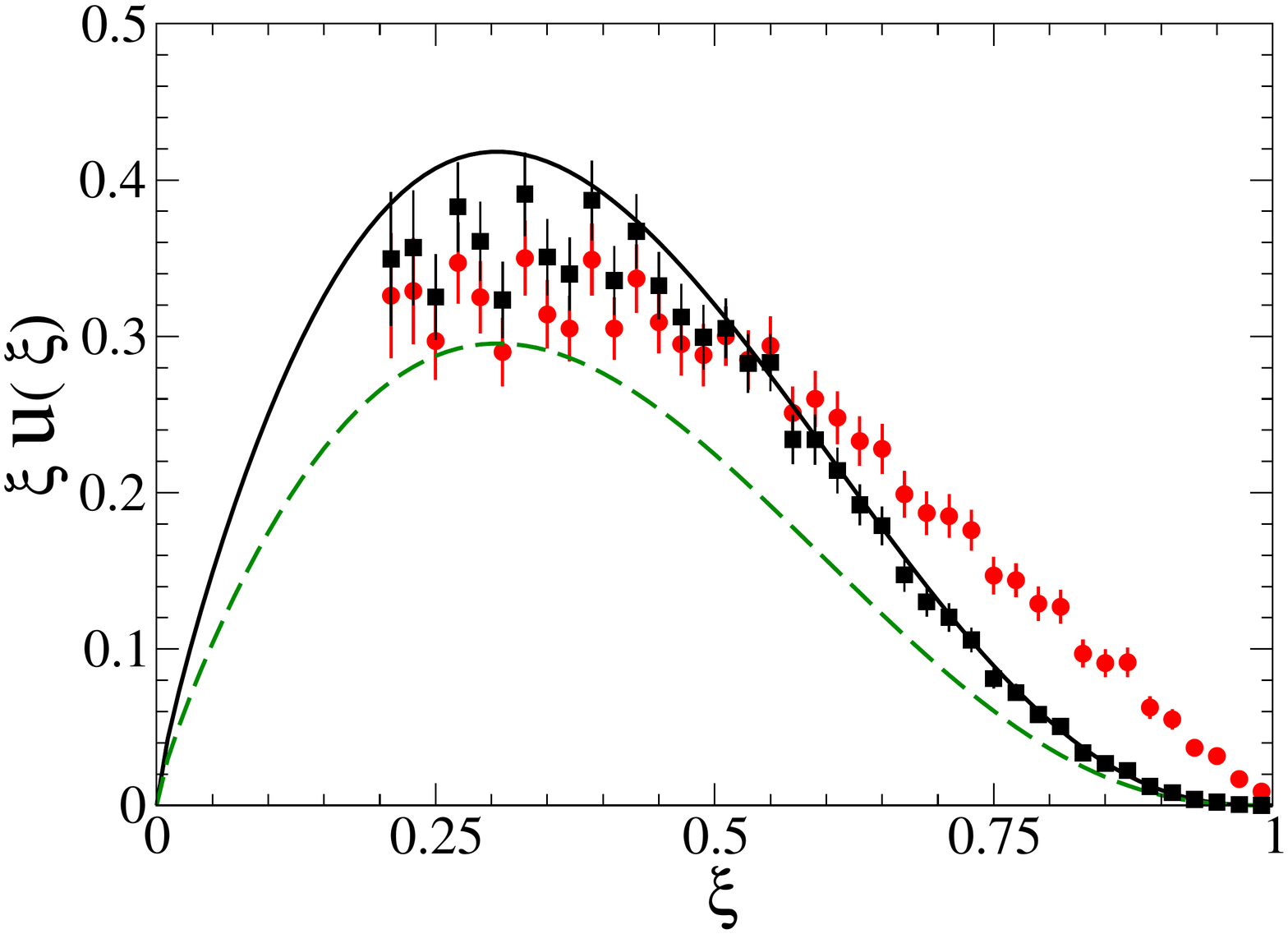,width=9.5cm}
\caption{(Color online). The distribution function $\xi~u(\xi)$ in a pion.  Solid line: full calculation (see Eqs. \eqref{Eq:pdf} and \eqref{Eq:uTMD1}), obtained from the BS amplitude solution of the BSE with  $m=255$ MeV,  $\mu= 637.5$ MeV and  $\Lambda=306$ MeV, and  evolved from the initial scale $Q_0=0.360$~GeV to $Q=5.2$~GeV (see text).
Dashed line: the evolved LF valence component, Eq. \eqref{Eq:uval}. Full dots: experimental data from Ref. \cite{ConwayPRD1989}. Full squares: reanalyzed data by using the ratio between the fit 3 of Ref. \cite{Aicher_PRL}, evolved to $5.2$ GeV, and the experimental data~\cite{ConwayPRD1989}, at each data point, so that the resummation effects (see text) are accounted for.
}\label{Fig2}
 \end{center}
\end{figure}
An ECLO evolution,
  as given in Ref. \cite{Cui:2021mom},   has been applied to the PDFs in Fig. \ref{Fig1} in order to  compare our results   to the E615 data \cite{ConwayPRD1989} (measured in Drell-Yan processes), and  { also taking} into account the reanalysis  carried out  in both
 Ref.~\cite{Wijesooriya:2005ir}, where  the scale $4.0$~GeV of the original experimental data  was suggested to be moved to  $Q=5.2$~GeV, and Ref.~\cite{Aicher_PRL}, where resummation effects
on the extraction of the pion PDF  were proposed. In particular,
 Fig.~\ref{Fig2} shows the comparison between { i) the theoretical calculations, full PDF and     valence contribution,  evolved  to $Q=5.2$~GeV,  ii)  the data originally delivered by the E615 Collaboration (assigned scale $4.0$~GeV)  and iii) the experimental data rescaled, at each $\xi$, by the  ratio between the fit 3 in  Ref.~\cite{Aicher_PRL}, properly evolved to $5.2$~GeV, and the E615 experimental data. Noteworthy, the calculations in Ref.~\cite{Aicher_PRL} have illustrated at which extent  the PDF extraction  from the experimental measurements  is  affected by the  resummation of the large logarithmic contributions  in the partonic hard-scattering cross sections.} It should be pointed out that the behavior of the evolved $\xi~u(\xi)$  for $\xi \to 1$ is given by $(1-\xi)^{\eta_5}$ { with  $\eta_5=2.94$ (with $\xi\in[0.9,1]$),}  to be compared, e.g., to the value $2.20\pm 0.64$ obtained by using recent  LQCD calculations \cite{Alexandrou:2021mmi}, where the PDF is reconstructed via Mellin moments, { as well as the exponent $2.81\pm 0.08$ reported  in Ref. \cite{Cui:2021mom}.
  The low-order Mellin moments for two   scales, $Q=2.0$ GeV and $5.2$ GeV, obtained from    our pion PDF (after properly evolving through ECLO) and  from the most recent LQCD results (with  $m_\pi=260$ MeV) ~\cite{Alexandrou:2020gxs,Alexandrou:2021mmi} are presented in Table~\ref{Tab1}.     }

\begin{table}[tbh]
 \caption{Low-order Mellin moments at two scales $Q=2.0$ GeV and $5.2$ GeV. First column: results  from our BSE approach, i.e. by using  Eqs. \eqref{Eq:pdf} and \eqref{Eq:uTMD1}  evolved  to  the scale $Q=2.0$ GeV from an initial scale $Q_0=0.360$ GeV (see text). Second column: LQCD calculations at $Q=2.0$ GeV from Ref. \cite{Alexandrou:2020gxs} for  $\langle x \rangle$, and Ref. \cite{Alexandrou:2021mmi} for   $\langle x^2 \rangle$ and  $\langle x^3 \rangle$, with the errors given by  statistical and systematic uncertainties added in quadrature. Third column: our results evolved to $5.2$ GeV. Fourth column: lattice result, evaluated by using the reconstructed PDF, as given  in Table XI of Ref. \cite{Alexandrou:2021mmi}.}
    \label{Tab1}
    \begin{center}
    \begin{tabular}{ccccc}
    \hline
    \hline
    ~~~& ~~BSE$_2$~~&LQCD$_2$ & BSE$_5$~~&LQCD$_5$ \\
    \hline
       $~~\langle x \rangle~~~$ &0.259  & 0.261$\pm$0.007& 0.221 & 0.229$\pm$0.008\\
        $\langle x^2 \rangle$&0.105 & 0.110$\pm$0.014 & 0.082 & 0.087$\pm$0.009\\
        $\langle x^3\rangle$& 0.052 &  0.024$\pm$0.018 & 0.039 & 0.042$\pm$0.010\\
        $\langle x^4\rangle$& 0.029&~ & 0.021 & 0.023$\pm$0.009 \\
        $\langle x^5\rangle$& 0.018&~& 0.012 & 0.014$\pm$0.007\\
        $\langle x^6\rangle$& 0.012&~& 0.008 & 0.009$\pm$0.005 \\
       \hline
    \end{tabular}
   
    \end{center}
\end{table}
Finally, in Fig.~\ref{Fig3}, the comparison is carried out with some recent theoretical outcomes obtained from different frameworks. The so-called continuum-QCD, based on the Dyson-Schwinger equation and the BSE, is able to yield the PDF, via Mellin moments evaluated in Euclidean space.
In particular, we compare with the results presented i) in Ref. \cite{Cui:2021mom},   where the PDF is obtained from  the  leading-twist two-particle distribution amplitude  (actual calculations are from Ref. \cite{LeiPRL13}), that depends upon only the anti-aligned component of the BS amplitude, and  ii)  in Ref. \cite{Bednar:2018mtf}, where the quark vertex is properly dressed (cfr. the bare vertex $\gamma^+$ in  Eq. \eqref{Eq:uTMD1}).
Moreover, it is shown the PDF directly calculated {within the BLFQ   in  Ref. \cite{Lan:2021wok,Lan:2022phd}}, and  the PDF  recently evaluated  in LQCD by the ETM Collaboration~\cite{Alexandrou:2021mmi}, by using the Mellin moments for the reconstruction. Interestingly,
the JAM NLO global fit analysis~\cite{Barry:2018ort} largely overlaps the original E615 data~\cite{ConwayPRD1989}. { It is rewarding that our dynamical calculation in Minkowski space  falls in the LQCD band and nicely agrees on the tail with { the recent calculations from both the DSE approach   \cite{Cui:2021mom} and the BLFQ one \cite{Lan:2021wok,Lan:2022phd}.}
\begin{figure}[htb]
\begin{center}
\epsfig{figure=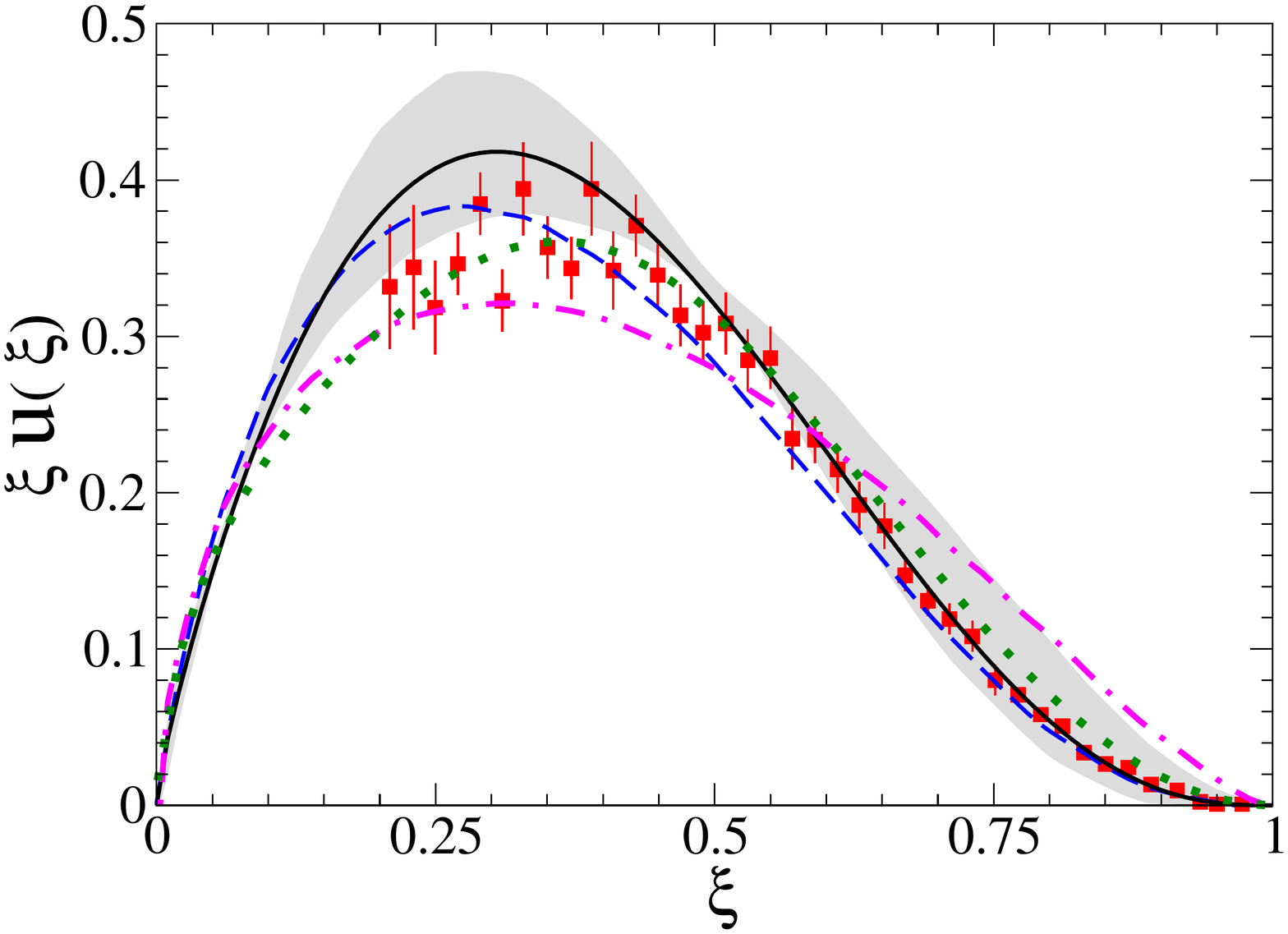,width=8.0cm}
\caption{(Color online). The distribution function $\xi~u(\xi)$ in a pion, at scale $5.2$ GeV.  Solid line:  full calculation as in Fig.~\ref{Fig2}.
Dashed line: DSE calculation from { Fig. 5} of Ref.~ \cite{Cui:2021mom}. Dash-dotted line: DSE calculation with dressed quark-photon vertex from Ref.~\cite{Bednar:2018mtf}.
Dotted line: { BLFQ result  ~\cite{Lan:2021wok,Lan:2022phd} }. Shaded area: LQCD calculation extracted via Mellin moments from Ref.~\cite{Alexandrou:2021mmi}.
}\label{Fig3}
 \end{center}
\end{figure}

{\it Summary.} For the first time, the pion PDF has been calculated  directly in Minkowski space,
within  a dynamical framework based on the 4D  Bethe-Salpeter equation. We have adopted the Nakanishi integral representation of the BS amplitude, so that the analytic behavior of the BS amplitude can be exposed and manipulated,  formally obtaining an equation for the so-called Nakanishi weight functions from the BSE.
 At this stage of development of our technology, only three parameters enter in the ladder kernel, namely:   the masses of quarks and gluon,
and the scale of the extended quark-gluon vertex.  The initial scale of our calculation is fixed at $Q_0=0.360$~GeV (cfr.  Ref.~\cite{Cui:2021mom}). The comparison with data and theoretical calculations are very encouraging and strongly motivates  improvements of our approach. In fact, we are currently working on including
 a consistent treatment of  quark and gluon self-energies (see, e.g., Refs. \cite{Jia:2019kbj,Mezrag:2020iuo}).

{\it Acknowledgments.}
 E. Y. gratefully thanks
INFN Sezione di Roma
 for providing the computer resources to perform all the calculations shown in this work. W. d. P. acknowledges the support from CNPq Grants No. 438562/2018-6 and
No. 313030/2021-9, and  CAPES
Grant No. 88881.309870/2018-01. T. F. acknowledges
the support from
CNPq (Grant No. 308486/2015-3) and FAPESP (Grants
No. 17/05660-0 and No. 2019/07767-1). J. H. A. N. acknowledges the support
from FAPESP Grant No. 2014/19094-8. E.
Y. acknowledges the support from FAPESP Grant No. 2016/25143 and No. 2018/21758-2. This work is a part of the
project Instituto Nacional de Ci\^encia e Tecnologia - F\'isica
Nuclear e Aplica\c c\~oes Proc. No. 464898/2014-5.


%

\end{document}